\documentclass[journal]{IEEEtran}

\ifCLASSINFOpdf
\else
\fi
\usepackage{graphicx}
\usepackage{subfig}
\usepackage{algorithm}
\usepackage{url}
\usepackage{algorithmic}
\usepackage{multirow}
\usepackage{bm}
\usepackage{amssymb}
\usepackage{amsmath}
\begin{document}
%
\title{Task-driven Semantic Coding via Reinforcement Learning}
%
%
%

\author{Xin Li, Jun Shi and
	Zhibo Chen,~\IEEEmembership{Senior Member,~IEEE,}
\thanks{Xin Li, Jun Shi, and Zhibo Chen are with the Department of Electronic Engineer and Information Science, University of Science and Technology of China, Hefei, Anhui, 230026, China (e-mail: lixin666@mail.ustc.edu.cn; shi1995@mail.ustc.edu.cn;
	chenzhibo@ustc.edu.cn). Xin Li and Jun Shi contribute equally to this paper. Corresponding author: Zhibo Chen.} 
}

\maketitle

\begin{abstract}
Task-driven semantic video/image coding has drawn considerable attention with the development of intelligent media applications, such as license plate detection, face detection, and medical diagnosis, which focuses on maintaining the semantic information of videos/images. Deep neural network (DNN)-based codecs have been studied for this purpose due to their inherent end-to-end optimization mechanism. However, the traditional hybrid coding framework cannot be optimized in an end-to-end manner, which makes task-driven semantic fidelity metric unable to be automatically integrated into the rate-distortion optimization process.
Therefore, it is still attractive and challenging to implement task-driven semantic coding with the traditional hybrid coding framework, which should still be widely used in practical industry for a long time.
To solve this challenge, we design semantic maps for different tasks to extract the pixelwise semantic fidelity for videos/images. Instead of directly integrating the semantic fidelity metric into traditional hybrid coding framework, we implement task-driven semantic coding by implementing semantic bit allocation based on reinforcement learning (RL). We formulate the semantic bit allocation problem as a Markov decision process (MDP) and utilize one RL agent to automatically determine the quantization parameters (QPs) for different coding units (CUs) according to the task-driven semantic fidelity metric. 
Extensive experiments on different tasks, such as classification, detection and segmentation, have demonstrated the superior performance of our approach by achieving an average bitrate saving of 34.39\% to 52.62\%  over the High Efficiency Video Coding (H.265/HEVC) anchor under equivalent task-related semantic fidelity.

\end{abstract}


\IEEEpeerreviewmaketitle

\section{Introduction}

\IEEEPARstart{W}{ith} the development of image/video analysis and understanding, many intelligent media applications, such as detection \cite{girshick2014rich, girshick2015fast, ren2015faster, redmon2016you, liu2016ssd,  hariharan2014simultaneous}, classification \cite{krizhevsky2012imagenet, simonyan2014very, he2016deep, huang2017densely, zoph2018learning}, retrieval \cite{saritha2019content, ng2020solar} and person reidentification \cite{zhang2020multi, jin2019semantics, jin2020style}, have been greatly promoted. These factors bring out the requirements for efficient compression of image/video signals, which can reduce the bitrate as much as possible when ensuring semantic fidelity for intelligent media applications. However, it is difficult to integrate the semantic distortion metrics directly into the traditional hybrid coding framework since the traditional hybrid coding framework cannot be optimized in an end-to-end manner. The current video coding standards are all based on the hybrid coding framework, such as Advanced Video Coding (H.264/AVC) \cite{wiegand2003overview}, High Efficiency Video Coding (H.265/HEVC) \cite{sullivan2012overview} and the recently released Versatile Video Coding (H.266/VVC) \cite{vtm}. They are still widely used in the industry, and should not be replaced by learning based video coding schemes in a short time. Therefore, it is necessary to explore an efficient task-driven semantic coding method for the traditional hybrid coding framework.

In the past few years, traditional image/video coding technologies have been devoted to improving the rate-distortion performance, such as Joint Photographic Experts Group (JPEG) \cite{wallace1992jpeg}, Better Portable Graphics (BPG), H.264/AVC, H.265/HEVC, and H.266/VVC. The distortion is usually measured by Mean Square Error (MSE) \cite{wang2009mean}, which can represent the pixel fidelity of an image/video. However, pixel fidelity cannot fully reflect the human perceptual viewing experience \cite{wang2009mean}. Therefore, many perceptual distortion metrics, such as structural similarity index (SSIM) \cite{wang2004image} and multi scale structural similarity index (MS-SSIM) \cite{wang2003multiscale}, have been proposed. To integrate perceptual distortions into traditional hybrid coding framework, the work \cite{ou2011ssim} implemented perceptual coding by changing the rate-distortion optimization with SSIM.
With the development of saliency detection \cite{zhang2017learning, wang2016saliency}, many perceptual coding schemes \cite{gao2016phase, li2017closed, ku2019bit}, which implemented bit allocation based on saliency detection, have been proposed.

Unlike pixel fidelity and perceptual fidelity metrics, semantic fidelity metrics are difficult to integrate into traditional coding frameworks since the traditional hybrid coding framework cannot be optimized in an end-to-end manner.
To solve this problem, a simple method is to build up an optimization schemes by heuristically adjusting image compression parameters (e.g., QP (quantization parameters)), such as \cite{liu2017recognizable}. However, this scheme cannot adaptively and automatically optimize coding configurations according to different semantic distortion metrics of different tasks. 
Recently, Bichon et al. \cite{bichon2019optimal} utilizes the psycho-vision guided function to weight the distortion in HEVC, which improves the subjective quality of encoded images. However, the weighting function for semantic coding is difficult to get through simple calculations. It’s still a considerable challenge to implement task-driven semantic coding with the traditional hybrid coding framework. 

In this paper,  we attempt to achieve task-driven semantic coding within traditional codecs by solving two essential problems, which are ``how to measure pixelwise semantic fidelity for different tasks" and ``how to integrate the semantic fidelity metric into the in-loop optimization of semantic coding". To solve the first problem, we utilize task-driven semantic maps, which are generated by Grad-CAM \cite{selvaraju2017grad} and Mask R-CNN \cite{he2017mask} to represent the pixelwise semantic importance of video/image. The semantic fidelity metric can be computed with the difference between the semantic maps before and after coding. The effectiveness of task-driven semantic maps can be seen in section \ref{semantic_map}. 
For the second problem, we implement semantic coding by formulating a semantic bit allocation scheme, i.e., deciding the quantization parameters of each coding unit(CU) as a Markovian decision process (MDP). Then, we introduce one RL agent to adaptively decide the quantization parameter (QP) for each CU by balancing the bit cost and semantic fidelity metric. According to RL-based semantic bit allocation, we can integrate the semantic fidelity metric into the in-loop optimization of semantic coding. When the training of the RL agent is completed, the whole process of the QP decision is off-policy, which can be processed in parallel with the encoder.

Since H.265/HEVC is the latest widely used video coding standard, we validate our task-driven semantic coding algorithm on H.265/HEVC in this paper, which is also easily generalized to other hybrid codecs such as H.264/AVC and H.266/VVC. To train the RL agent efficiently, we built a universal task-driven semantic coding dataset.  
In this dataset, the semantic importance maps are generated by Grad-CAM for classification and Mask R-CNN for segementation and detection. The bit costs are extracted from the traditional coding framework H.265/HEVC \cite{sullivan2012overview}. 
With this dataset, the RL agent can be guided to make precise decision for task-driven semantic coding. Extensive experiments have demonstrated that our scheme can achieve an average bitrate reduction from 32.4\% to 52.6\% with comparable task-driven semantic fidelity.

The main contributions of our work presented in this paper can be summarized as follows:
\begin{itemize}
	\item To the best of our knowledge, we are the first to implement task-driven semantic coding for the traditional hybrid coding framework by adaptive semantic bit allocation with reinforcement learning.
	
	\item We succeed in measuring the task-related pixelwise semantic fidelity with semantic importance map differences and integrate the semantic fidelity metric into the in-loop optimization of semantic coding.
	
	\item We create a dataset\footnote{The dataset and code will be released at \url{http://staff.ustc.edu.cn/~chenzhibo/resources.html}} that is suitable for the RL agent to learn to decide quantization parameters. Extensive experiments on various intelligent tasks validate that our algorithm can achieve the average bitrate reduction from 32.4\% to 52.6\% with comparable task-driven semantic fidelity.
\end{itemize}

The work described in this paper is related our previous work that was reported in \cite{shi2020reinforced}. In our previous paper, we presented the basic idea of RL-based semantic bit allocation and provide preliminary results. 
The difference of this paper and \cite{shi2020reinforced} are presented as follows. First, in this paper, we built a complete and general task-driven semantic coding framework by introducing the pixel-level semantic map, which can unify the different tasks and improve the scalability and generalization of our semantic coding scheme. Second, we validate the effectiveness of our pixelwise semantic map to represent the task-driven semantic fidelity. Third, we make a more reasonable evaluation for our task-driven semantic coding scheme by conducting the performance evaluation in  full bitrate (from low bitrate to high bitrate) space. Finally, more thoroughly experiment results and ablation studies are provided in this paper verify the effectiveness of the proposed framework.


The rest of the paper is organized as follows. In Section II, we briefly review related works of image/video compression and intelligent tasks. Section III describes our  task-driven semantic coding scheme via reinforcement learning in detail. This algorithm is named RL-based semantic coding (RSC). Section IV introduces our dataset generation together with extensive experiments and ablation studies. Finally, we draw conclusions and provide directions for future work in Section V.

\section{Related Work}
\subsection{Task related Coding}
In recent years, the traditional hybrid coding framework has not been able to satisfy people’s needs in some situations. Therefore, many task related coding schemes, such as perceptual coding and semantic coding, have been explored \cite{torfason2018towards, huang2010perceptual, hadizadeh2011saliency, chen2019learning, zhou2020just} with the development of image/video coding techniques.

To improve the perceptual quality of image/video, Huang et al. \cite{huang2010perceptual} first proposed developing a method of perceptual rate-distortion optimization by applying SSIM as a quality metric in H.264, which can improve the overall perceptual quality of encoded images/videos. However, it is unable to adapt to improve the perceptual quality of regions in which people are interested. With the advance of saliency detection, some researchers have succeeded in detecting the regions of interest in images/videos. Based on saliency detection, some works \cite{guo2009novel, li2011visual, khanna2015perceptual, hadizadeh2011saliency} succeed in improving the subjective quality of coded video by implementing visual attention guided bit allocation in video compression.
In recent years, learning-based perceptual coding \cite{ki2018learning, cheng2019perceptual} has been explored to further improve the perceptual quality of encoded images/videos.

With the development of deep learning, some deep neural network (DNN)-based image/video coding frameworks that focus on semantic information of image/video have been proposed. Torfason et al. \cite{torfason2018towards} let the compressed representation reserve the semantic information by jointly training compression networks with image understanding tasks on the compressed representations. Furthermore, Luo et al.  \cite{luo2018deepsic} proposed a deep semantic image compression (DeepSIC) model that enables the compressed code stream to carry semantic information of the image during its storage and transmission by pre-semantic analysis or post-semantic analysis. Unlike transmitting semantic information through compressed representations without decoding, Akbari et al. \cite{akbari2019dsslic} decomposed images into thumbnails and segmentation maps. Then they transmitted the semantic information by coding the segmentation maps and reconstructed the decoded images by utilizing the semantic information, which achieves more bit saving as well by transmitting the segmentation maps. To improve the compression quality of facial compression, Chen et al. \cite{chen2019learning} integrated the semantic fidelity into a facial compression framework, which employed a generative adversarial network (GAN) as a metric. However, the traditional hybrid coding framework cannot be optimized in an end-to-end manner, which has been used in the above works. It is still attractive and challenging to implement task-driven semantic coding with the traditional hybrid coding framework. In this paper, we achieve task-driven semantic coding for traditional hybrid coding framework by achieving semantic bit allocation for HEVC intra coding based on reinforcement learning (RL).

\subsection{Related computer vision tasks}
Deep learning has revolutionized many computer vision fields, especially in image/video understanding and analysis. Many intelligent applications, such as object detection, segmentation, classification, and face recognition, have been greatly developed. Here, we briefly introduce the classification, segmentation and detection.

Since AlexNet \cite{krizhevsky2012imagenet} won the 2012 ImageNet LSVRC championship, many deep convolution neural networks have been proposed to improve the classification accuracy, including VGGNet \cite{simonyan2014very}, InceptionNet \cite{chen2017rethinking}, ResNet \cite{he2016deep}, DenseNet \cite{huang2017densely} and NASNet \cite{zoph2018learning}. These networks have been applied to various computer vision tasks as backbones, including pose estimation, person re-identification and image restoration, due to their powerful feature extraction capability. 
\begin{figure*}[htp]
	\begin{center}
		\includegraphics[width=0.95\linewidth]{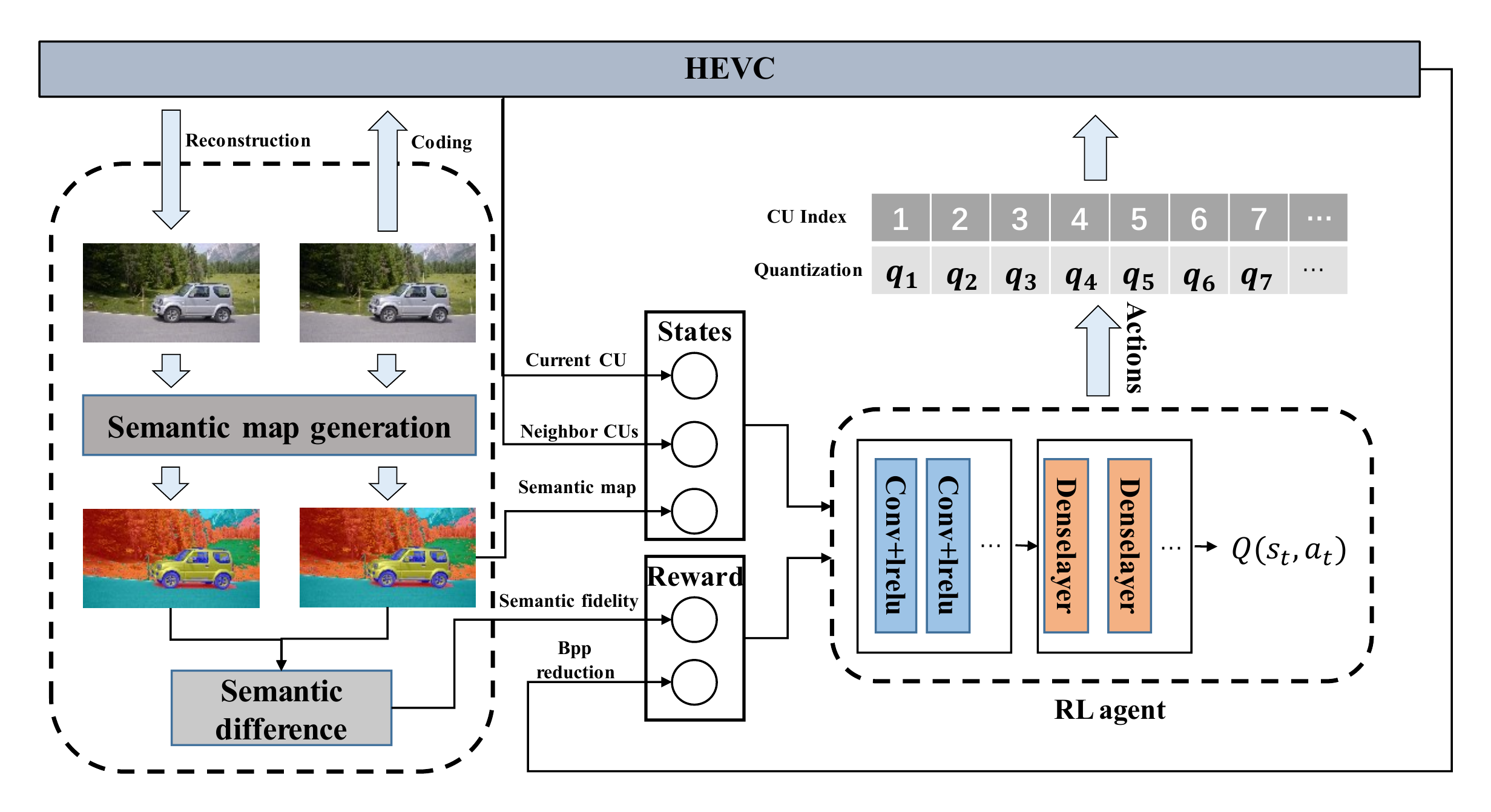}	
	\end{center}
	\caption{Illustration of our task-driven semantic coding via reinforcement learning. It is achieved by semantic bit allocation with H.265/HEVC. Given an original image and the task, the image first passes the semantic map generation module to generate the task-driven semantic map. Then, the agent takes an action based on the observation, including the coding unit, its neighbor coding units and the semantic map. In other words, the agent selects the corresponding quantization parameters for the coding unit of the original image. Next, according to the semantic difference between the original image with the reconstructed image and bit reduction, we calculate the score as the reward to update the agent, which balances semantic fidelity and bitrate. After training, the agent can implement task-driven semantic coding with H.265/HEVC.}	
	\label{fig:framework}
\end{figure*}
Additionally, with category-independent region proposals and supervised pretraining for auxiliary tasks, R-CNN \cite{girshick2014rich} made a breakthrough in the object detection field. However, the training and object detection process of R-CNN is expensive in space and time. Thus, Fast R-CNN \cite{girshick2015fast} applied SPPnet \cite{he2015spatial} to speed up R-CNN by sharing a feature map instead of performing a ConvNet forward pass for each object proposal. Faster R-CNN \cite{ren2015faster} introduced a novel region proposal network (RPN) that shares full-image convolutional features with the detection network where proposal computation is nearly cost-free. Instead of adapting the mechanism of region proposals, Redmon et al. proposed a new framework named YOLO \cite{redmon2016you}, which redefines object detection as a regression problem and separates bounding-boxes regression and associated class probabilities. However, it cannot locate the object accurately. Thus, SSD \cite{liu2016ssd} solved the problem by combining the compression ideas of YOLO and the anchor mechanism of Faster R-CNN.

Unlike the object detection task, which returns the coordinates of the bounding-box, segmentation aims to label every pixel in an image with its class. Early works \cite{hariharan2014simultaneous,hariharan2015hypercolumns,dai2015convolutional,pinheiro2015learning} focused on noninstance semantic segmentation, which is mainly based on bottom-up segments. However, the upper layers cannot capture rich spatial information. To refine the coarse  object segments, \cite{pinheiro2015learning} proposed a novel bottom-up/top-down architecture that combines rich spatial information and object-level knowledge to obtain more accurate segmentation. Based on the development of semantic segmentation, many instance-aware semantic segmentations \cite{dai2016instance,li2017fully,kirillov2017instancecut,bai2017deep,arnab2017pixelwise} have been proposed that can label pixels according to not only their class but also the object to which they belong. Recently, \cite{he2017mask} is the state-of-the-art work in instance segmentation, bounding-box object detection with ROIAlign and predicting an object mask in parallel with bounding-box recognition. Furthermore, Huang et al. achieved more improvement with a mask scoring strategy based on Mask-RCNN \cite{huang2019mask}. In this paper, we employ the Mask-RCNN to validate the effectiveness of our algorithm on the tasks of segmentation and detection. 

\section{RL-based semantic coding (RSC)}
In this section, we introduce our task-driven semantic coding algorithm RSC. The essence of task-driven semantic coding is to balance the semantic fidelity and coding cost. As shown in Fig. \ref{fig:framework}, we achieve task-driven semantic coding by RL-based semantic bit allocation with H.265/HEVC. The overall architecture mainly contains two parts: task-driven semantic map generation and the RL agent for semantic bit allocation. Therefore, we first formulate the semantic bit allocation based on traditional MSE-based bit allocation. Then we detail the components  of RL for semantic bit allocation, including the state, action, reward and agent architecture.

\subsection{Semantic Bit Allocation}
Bit allocation can usually be implemented on three levels: GOP level, frame level and basic coding unit level. In this article, we mainly focus on coding unit (CU)-level semantic bit allocation.

In the process of bit allocation, we usually minimize the distortion $D$ with a given number of bits ${{R}_{c}}$, which is formulated by:
\begin{equation}
\label{equation_1}
\min \sum{{{D}_{i}},s.t.\sum{{{r}_{i}}\le {{R}_{c}}}},
\end{equation}
where ${{R}_{c}}$ is the upper limit of the coding bits. ${{D}_{i}}$ and ${{R}_{i}}$ are the distortion and coding bits for the $i$th basic unit. This constrained optimization problem can be converted to an unconstrained optimization problem using the Lagrange optimization method as follows:
\begin{equation}
\label{equation_2}
\min J,J=\sum{{{D}_{i}}\text{+}\lambda \sum{{{R}_{i}}}}.
\end{equation}
Here, $J$ represents rate-distortion performance which is one of fundamental considerations in bit allocation. Then, we can obtain the optimal solution to formula \ref{equation_2} by taking the derivative of formula \ref{equation_2} as:
\begin{equation}
\label{equation_3}
\lambda \text{=-}\frac{\partial D}{\partial R},s.t.D=\sum{{{D}_{i}},R=\sum{{{R}_{i}}}},
\end{equation}
where $D$ and $R$ represent the total distortion and total bits, respectively, for encoding one coding tree Unit(CTU).
To solve the formula \ref{equation_3}, Dai  et al. \cite{dai2004rate} modeled the relationship of $R$ and $D$ with the hyperbolic function when the distortion is MSE as:
\begin{equation}
\label{equation_4}
D(R)=C{{R}^{-K}},
\end{equation}
where $C$ and $K$ are model parameters, which are determined by the characteristics of the source block. Then we can obtain the slop of the R-D curve $\lambda$ with formula \ref{equation_4} as:
\begin{equation}
\label{equation_5}
\lambda =CK{{R}^{-K-1}}\triangleq \alpha {{R}^{\beta }},
\end{equation}
and simultaneously we can get $R$ by transforming the formula \ref{equation_5} as:
\begin{equation}
\label{equation_6}
R={{\left( \frac{\lambda }{\alpha } \right)}^{\frac{1}{\beta }}}={{\alpha }_{1}}{{\lambda }^{{{\beta }_{1}}}}.
\end{equation}
From formula \ref{equation_6}, we can observe that the bits $R$ are determined by the parameter $\lambda$ for a certain block. Therefore, in the process of bit allocation which is based on the distortion MSE or SSIM, ${{\alpha}_{1}}$ and ${{\beta}_{1}}$ are usually computed by precoding the block. Then, we can perform bit allocation by computing $\lambda$ with a given $R$ for a coding block according to formula \ref{equation_5}. As shown in \cite{li2016optimal} and \cite{li2014lambda}, the best quantization parameter $estQP$ corresponding to coding parameter $\lambda$ can be computed by:
\begin{equation}
\label{equation_7}
\text{estQP}=4.2005\ln \lambda +13.7122.
\end{equation} 
However, the above bit allocation algorithm is designed for the distortion metric MSE, which cannot directly be applicable to  semantic coding since the semantic distortion metric is more complex without an empirical formula.  Therefore, we propose a new semantic bit allocation algorithm for semantic coding based on the above bit allocation algorithm in this section.

Based on formula \ref{equation_2}, the semantic bit allocation can be modeled as:
\begin{equation}
\label{equation_8}
\min {{J}_{s}},{{J}_{s}}=\sum{{{D}_{si}}}+{{\lambda }_{s}}\sum{{{R}_{i}}},
\end{equation}
where ${{J}_{s}}$ is semantic rate distortion for one frame. ${{D}_{si}}$ and ${{R}_{i}}$ represent the semantic distortion and encoding bits, respectively, of the $i$th block in one frame. And the definition of ${{D}_{si}}$ can be seen in equation \ref{equation_12} of section III. B.
${{\lambda }_{s}}$ is an adjustable semantic coding parameter that is responsible for balancing ${{D}_{si}}$ and ${{R}_{i}}$.  Unfortunately, we cannot obtain the optimal solution to formula \ref{equation_8} directly because there is no formula that can characterize the relationship between ${{D}_{si}}$ and ${{R}_{i}}$ such as MSE-based bit allocation. 
Considering that the semantic rate-distortion optimization is a Markov process, which decides to increase or decrease ${{R}_{i}}$ by observing the change in the results of the former state,  we employ reinforcement learning (RL) to  derive the optimal solution ${{estR}_{i}}$, and we express this process with:
\begin{equation}
\label{equation_9}
est{{R}_{i}}=RL(\min {{J}_{s}}).
\end{equation}
Actually, in the traditional hybrid coding framework, the encoding bits ${{R}_{i}}$ is adjusted by the quantization parameter (QP) directly. Therefore, in this paper, we utilize RL agent to determine the best quantization parameters as equation \ref{equation_10}.
\begin{equation}
	\label{equation_10}
	est{QP}=RL(\min {{J}_{s}}).
\end{equation}

\subsection{Semantic importance map generation}
In the learning based coding framework, the formula \ref{equation_8} can be optimized easily with end-to-end training, because the semantic distortion can be substituted with the feature difference. However, it is not possible to apply this strategy to the traditional hybrid coding framework since the traditional coding framework cannot be optimized in an end-to-end manner.
Thus, to optimize the formula \ref{equation_8}, the semantic distortion ${{D}_{si}}$ must be expressed explicitly. Thus, we have to convert abstract semantic information to a measurable form first. 

\begin{figure}[t]
\begin{center}
	\includegraphics[width=1.0\linewidth]{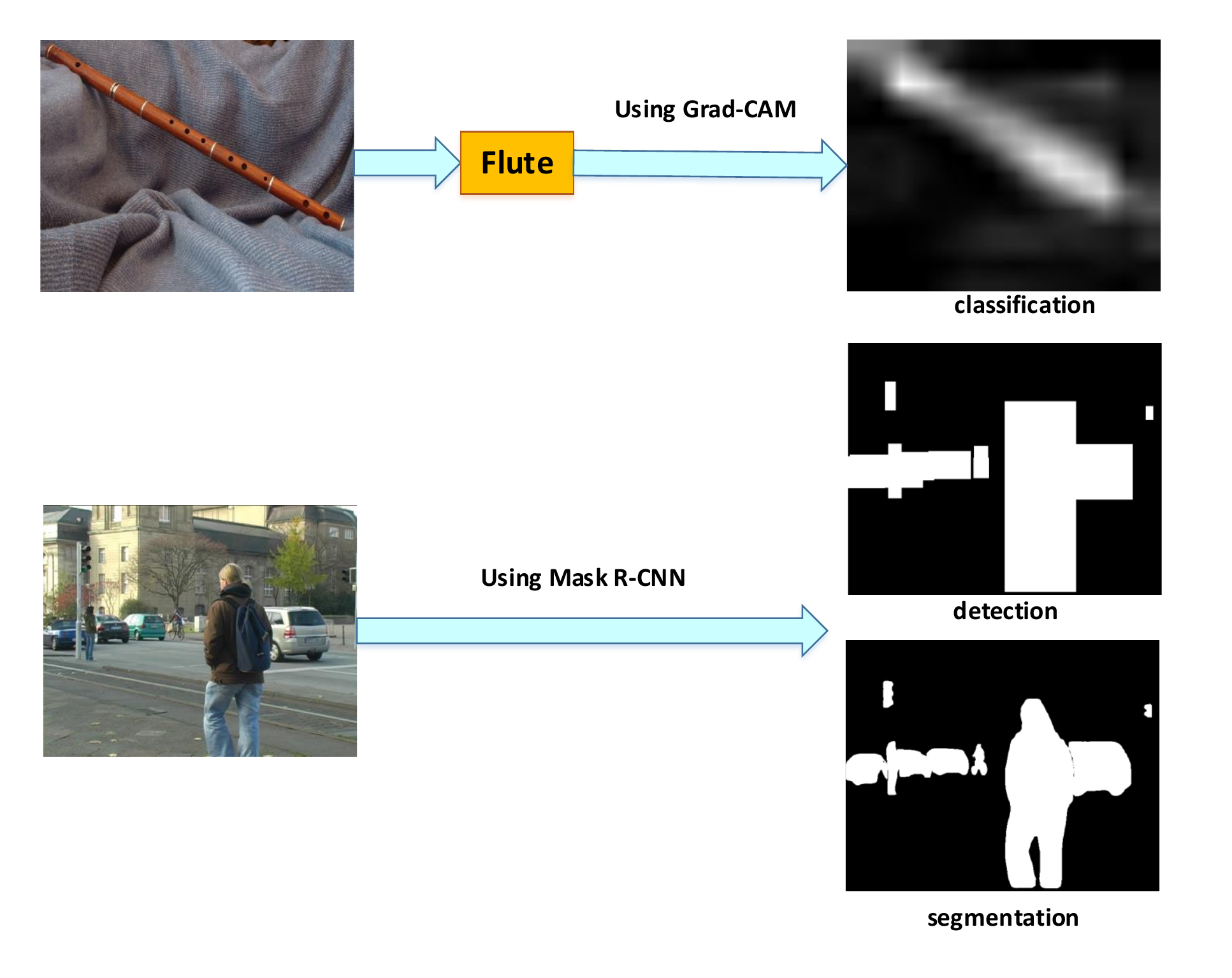}	
\end{center}
\caption{Semantic importance map generation}	
\label{fig:IMG}
\end{figure}

For some intelligent tasks such as classification, pose estimation, and person-reID, the outputs can accurately represent the semantic information that the tasks are concerned about. Thus, we can obtain the picture-level semantic importance map ${{M}_{s}}$ by extracting semantic information from the corresponding outputs. The process of task-driven semantic importance map generation is shown in Fig. \ref{fig:IMG}. 
\begin{figure*}[htp]
	\begin{center}
		\includegraphics[width=1.0\textwidth]{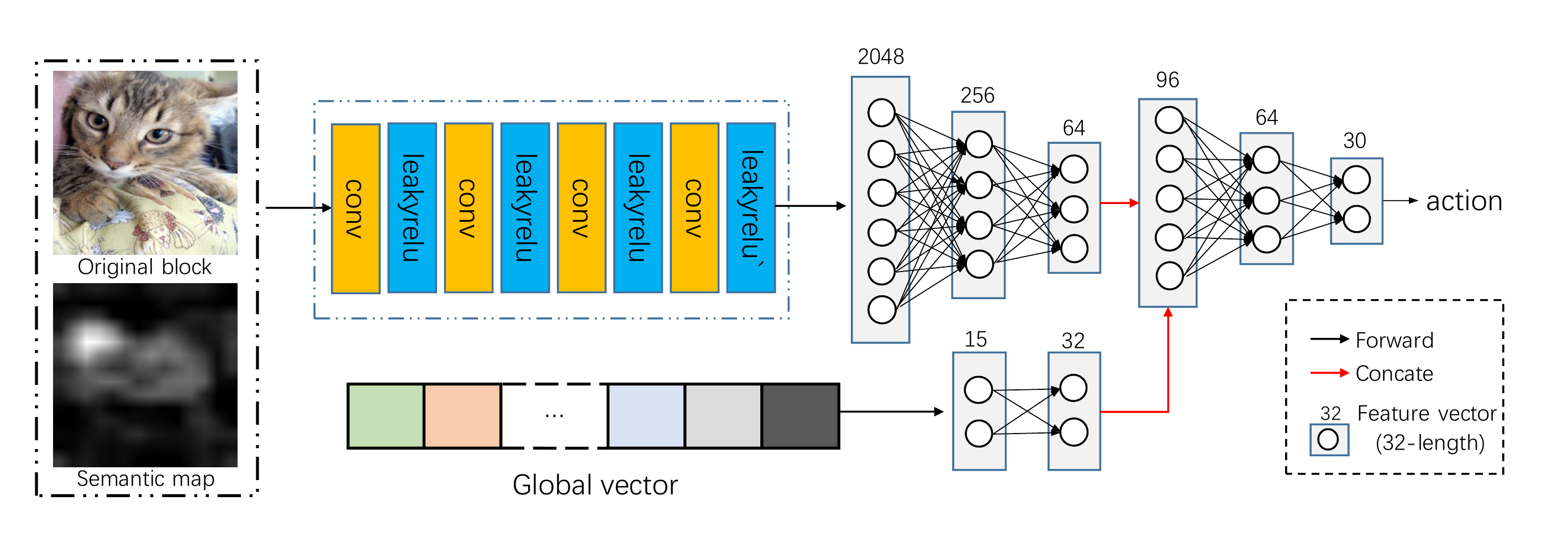}	
	\end{center}
	\caption{Structure of the proposed Q-network. There are two input branches: the current block part and the global information part. The luminance and importance map of current blocks are concatenated in depth to represent local information}	
	\label{fig:DQN}
\end{figure*}

As seen in Fig. \ref{fig:IMG}, the outputs of the task segmentation and detection can be mapped to the picture level easily. Thus, we employ Mask-RCNN to obtain outputs of detection and segmentation and directly use the outputs as task-driven semantic importance maps ${{M}_{s}}$. However, the results of the classification task cannot be directly mapped to ${{M}_{s}}$ because the output is only a label. Inspired by \cite{selvaraju2017grad}, we implement Grad-Cam, which uses the gradient backpropagation to flow into the final convolution layer of the CNN model to produce a localization map that highlights the semantic importance regions for predicting the concept. The specific CNN model we adopt is VGG-19. Then we can obtain the semantic importance maps of frames for the classification task by Grad-Cam.

With the task-driven semantic map, we succeed in converting abstract semantic information to a measurable form ${{M}_{s}}$. The amount of semantic information is indicated by the pixel value of ${{M}_{s}}$. Then we can measure the task-driven semantic distortion by:
\begin{equation}
\label{equation_12}
{{D}_{si}}=\Delta {{M}_{s}}.
\end{equation} 
Here, $\Delta {{M}_{s}}$ represents the difference in the semantic importance map before and after coding. And the difference is calculated with L1 normalization.
Finally, the RL agents can obtain the optimal solution to formula \ref{equation_8} with the measurable semantic distortion ${{D}_{si}}$.
\subsection{Reinforcement Learning for Semantic Bit Allocation}
After obtaining the ${{M}_{s}}$ of one frame, a simple bit allocation method is to heuristically increase the bitrate for the highly weighted area, such as the threshold scheme. However, these heuristic methods can hardly obtain the optimal results of formula \ref{equation_8} and may introduce limitations because the results rely heavily on the handcrafted designs. Recently, RL has achieved outstanding performance in many tasks, especially in unsupervised or semisupervised scenarios. It has also been used in many approaches \cite{li2019reinforcement, chen2018reinforcement, hu2018reinforcement, helle2017reinforcement} to optimize the traditional hybrid coding framework. In this paper, we adopt the reinforcement learning algorithm, Deep Q-Learning (DQN) \cite{mnih2013playing}, to solve the semantic bit allocation problem. 

The details are as follows:

\subsubsection{Problem formulation}
to obtain the optimal $est{{\lambda }_{i}}$ and corresponding $estQ{{P}_{si}}$ for each block, we have to obtain the optimal solution ${{R}_{i}}$ to the formula \ref{equation_8}. We formulate this process as an MDP process in this paper, which includes five elements: state, action, reward, state transition probability and policy(agent). Then, an agent is designed to observe the states from the environment, and executes a series of actions to optimize the formula \ref{equation_8}. In this process, the state transition probability ${{P}_{sa}}$ is 1 because the environment is deterministic. The remaining elements are detailed below:

\subsubsection{State}
 To determine the action for the next step, the agent must observe the states of the coding block and global information of the frame. Therefore, the coding block states are sent to the agent according to the encode order, $i.e.$ from left to right and from top to bottom. Here, the state of the coding block consists of the luminance, semantic importance map of the current block and a 15-d feature vector that can reflect the global information of the frame. The details of the feature vector are shown in Table. \ref{tab:global_info} .

\begin{table}[htp]
	\centering
	\caption{Components of global vector}
	\setlength{\tabcolsep}{7mm}{
 	\begin{tabular}{c|c}
 		\hline
 		Index & Vector components                  \\ \hline
 		1     & Number of overall CUs              \\ \hline
 		2     & Index of current CU                \\ \hline
 		3     & Mask ratio of current CU           \\ \hline
 		4-7   & Mask ratio of neighboring CUs      \\ \hline
 		8     & Mask ratio of overall frame        \\ \hline
 		9     & Instance number of current CU      \\ \hline
 		10-13 & Instance number of neighboring CUs \\ \hline
 		4-15  & QPs of left and above CUs          \\ \hline
 	\end{tabular}}
 \label{tab:global_info}
\end{table}

\subsubsection{Action}
To obtain optimal coding bits ${{estQP}_{i}}$ in formula \ref{equation_8}, we have to take the action to change the coding bits ${{R}_{i}}$. In traditional coding framework such as H.265/HEVC, the coding bits and coding quality usually change by directly adjusting the quantization Parameter (QP). Lower QP leads to a higher bitrate and less distortion. From formula \ref{equation_7}, we can find that the corresponding coding parameter $\lambda$ can be computed by:
\begin{equation}
\label{equation_13}
\lambda \text{=exp}\left\{ \frac{QP-13.7122}{4.2005} \right\}.
\end{equation}
Thus, to meet the above coding method and  reduce action space complexity, we take the action by optimally selecting the QP value. Specifically, the action space contains the QP values from 22 to 51. For one CTU, the agent can choose the best strategy to determine the optimal Quantization Parameter (QP) to encode this CTU according to the observation.

\subsubsection{Reward}   
As a method for evaluating the action, the cumulative reward is the optimization goal of MDP. The agent is guided to learn the optimal action by observing the reward. In this paper, the formula \ref{equation_8} is our optimization goal in the semantic bit allocation process. When we set the base action as 22, which can be adjusted according to different tasks, the corresponding rate distortion $J_{s}^{22}$ can be expressed by:
\begin{equation}
\label{equation_14}
J_{s}^{22}=\sum{D_{si}^{22}+{{\lambda }_{s}}}\sum{R_{i}^{22}},
\end{equation}
where $D_{s}^{22}$ and $R_{i}^{22}$ are the semantic distortion and coding bits, respectively, when coding the frame with QP 22. Then the convex optimization problem  formula \ref{equation_8} can be transformed to:
\begin{equation}
\label{equation_15}\
\begin{aligned}
& \max \Delta {{J}_{s}}, \\ 
& \Delta {{J}_{s}}=J_{s}^{22}-{{J}_{s}} \\ 
& =\sum{\left( D_{si}^{22}-{{D}_{si}} \right)}+{{\lambda }_{s}}\sum{\left( R_{i}^{22}-{{R}_{i}} \right)} \\ 
& =-\sum{\Delta M_{s}^{a}+{{\lambda }_{s}}\sum{\Delta {{R}_{i}}}} \\ 
& =-\sum{\Delta M_{s}^{a}+{{\lambda }_{s}}N\sum{\Delta Bp{{p}_{i}}}}
\end{aligned},
\end{equation}
where $\Delta M_{s}^{a}$, $\Delta {{R}_{i}}$ and $\Delta Bp{{p}_{i}}$ represent the difference in semantic information, coding bits and coding bits per pixel (bpps), respectively, between the coding block with QP 22 and the coding block after taking the action.	$N$ is the number of pixels in one coding block. This formula \ref{equation_15} is equivalent to:
\begin{equation}
\label{equation_16}
\max Reward={{\sum{\Delta Bpp}}_{i}}-{\alpha}_{s} \sum{\Delta M_{s}^{a}},{\alpha}_{s} =\frac{1}{{{\lambda }_{s}}N}.
\end{equation}
In the above formula, $Reward$ is the reward of the MDP, which instructs the agent to take the action. The $\alpha$ is an adjustable parameter that is related to the semantic coding parameters ${\lambda }_{s}$ and $N$. Therefore, we can optimize different semantic coding models by adjusting the parameter $\alpha$.

\subsubsection{Agent}
In this paper, the agent is used to predict the optimal action for every block with a Q-network. Taking the state ${{s}_{t}}$ as input, the Q-network outputs the decayed cumulative reward (Q-value) of each action $a$ as $Q\left( {{s}_{t}},a \right)$. Then, we can obtain the optimal action $a_{t}^{*}$ with:
\begin{equation}
\label{equation_17}
a_{t}^{*}=\underset{a}{\mathop{\arg \max }}\,Q({{s}_{t}},a).
\end{equation}
\begin{figure*}[htp]
	\begin{center}
		\includegraphics[width=1.0\textwidth]{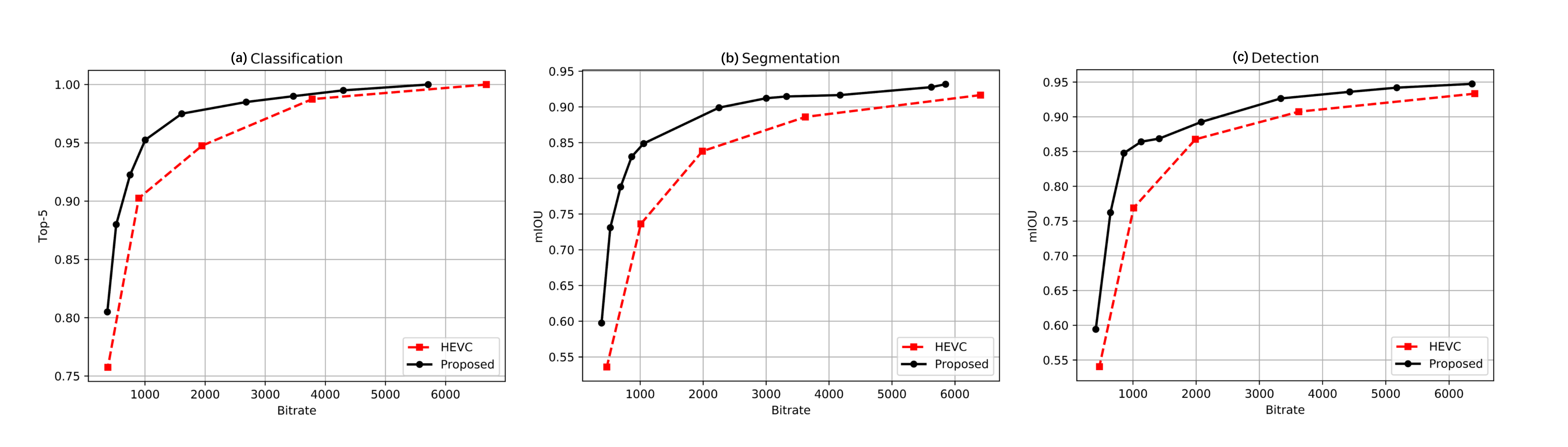}	
	\end{center}
	\caption{Comparisons of classification, segmentation and detection between the proposed RSC algorithm and H.265/HEVC. }	
	\label{fig:result}
\end{figure*}

For the Q-network structure, we have two input branches: the current block part and the global feature vector. The block information flows through four convolution layers and two concatenate the features with the global feature vector after ascending the dimension together. The combination of these features can help the agent to better understand the environment. Next, the overall features flow through two fully connected layers, including one hidden layer and one output layer. All convolution layers and hidden fully connected layers are activated with a leaky rectified linear unit with $\alpha \text{=}0.25$, while the output layer is not activated. Fig. \ref{fig:DQN} shows the details of the proposed Q-network.
\begin{algorithm}[htb]
	\renewcommand{\algorithmicrequire}{\textbf{Input:}}
	\renewcommand{\algorithmicensure}{\textbf{Output:}}
	\caption{Training process}
	\label{alg:Training}
	\begin{algorithmic}[1]
		\REQUIRE Original images ${I}_{0}$ and their semantic maps ${M}_{s}^{0}$
		\ENSURE Well-trained RL agent
		\STATE Initialize the network parameters $\theta$
		\FOR {Frame f = 0 to M-1}
		\FOR {CU t = 0 to N-1}
		\STATE Set state ${s}_{t}$
		\STATE Choose action $a_{t}=\underset{a}{\mathop{\arg \max }}\,Q({{s}_{t}},a)$
		\STATE Encode the CU with quantization parameter $QP=a_{t}+22$
		\STATE Obtain $bpp$ and corresponding ${M}_{s}$
		\STATE Compute reward ${r}_{t+1}$ as Equ. \ref{equation_16}
		\STATE Observe reward ${r}_{t+1}$ and next state ${s}_{t+1}$
		\STATE Store transition $({s}_{t}, {a}_{t}, {r}_{t+1}, {s}_{t+1})$
		\STATE Sample a mini-batch of transitions from buffer $({s}_{t'}, {a}_{t'}, {r}_{t'+1}, {s}_{t'+1})$
		\STATE Compute ${{y}_{{{t}'}}}={{r}_{{{t}'+1}}}+\gamma {{\max }_{{{a}_{{t}'+1}}}}Q({{s}_{{t}'+1}},{{a}_{{t}'+1}};\theta ) $
		\STATE Update $\theta$ with the loss ${{\sum\nolimits_{{{t}'}}{[{{y}_{{{t}'}}}-Q({{s}_{{{t}'}}},{{a}_{{{t}'}}};\theta )]}}^{2}} $
		\ENDFOR
		\ENDFOR 
	\end{algorithmic}
\end{algorithm} 

\begin{algorithm}[htb]
	\renewcommand{\algorithmicrequire}{\textbf{Input:}}
	\renewcommand{\algorithmicensure}{\textbf{Output:}}
	\caption{Process of RSC}
	\label{alg:process_RSC}
	\begin{algorithmic}[1]
		\REQUIRE Original image ${I}_{0}$
		\ENSURE Bit stream and reconstructed image ${I}_{c}$
		\STATE Generate the semantic map ${M}_{s}^{0}$ for ${I}_{0}$
		\STATE Obtain the QPs for each CU from the RL agent
		\STATE Input the QPs for each CU to the HM16.19 and compress the original image ${I}_{0}$
		\STATE Obtain the bit stream and reconstructed image ${I}_{c}$
	\end{algorithmic}
\end{algorithm} 

 \begin{figure*}[htp]
	\begin{center}
		\includegraphics[width=1.0\textwidth]{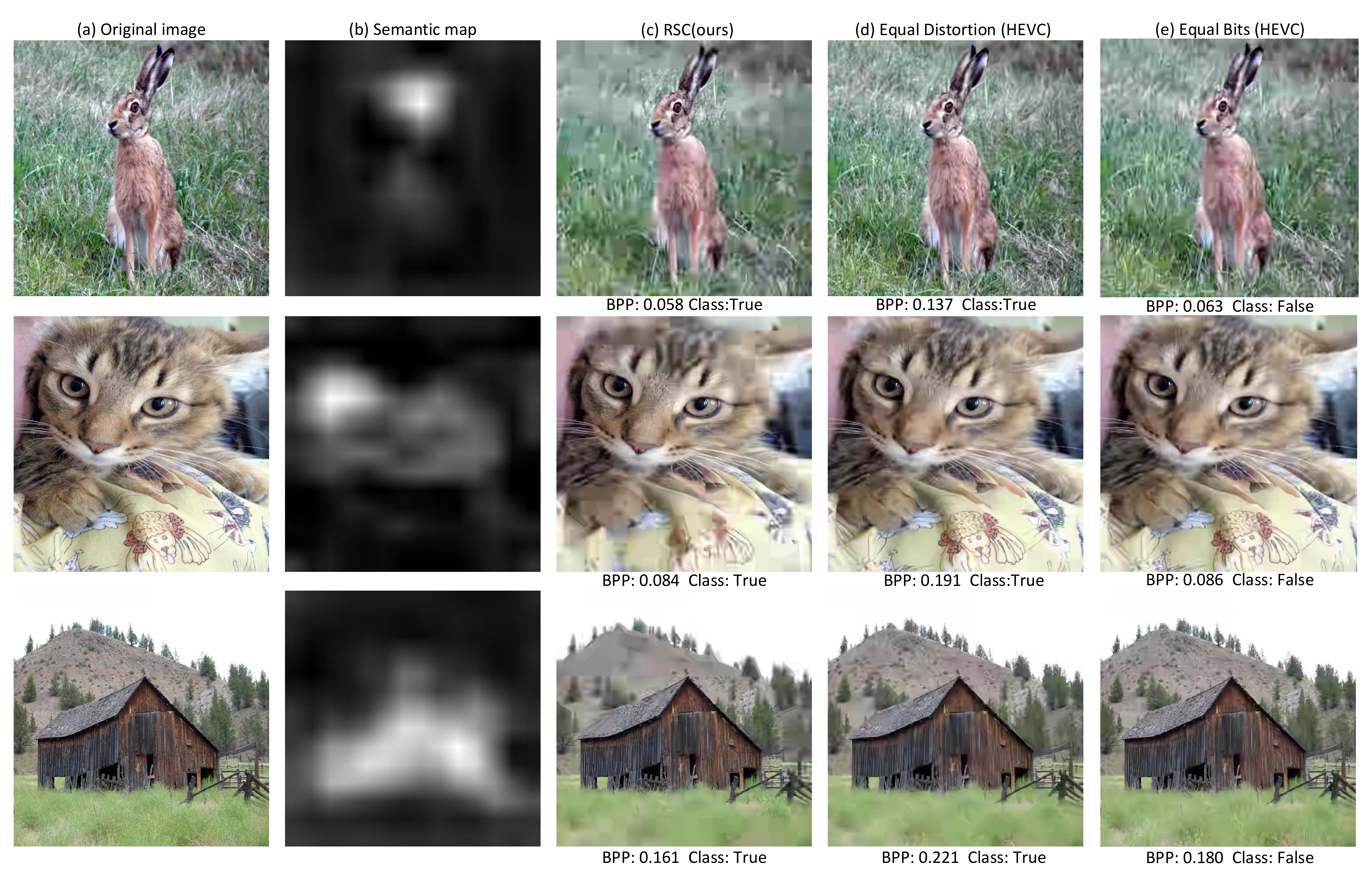}	
	\end{center}
	\caption{Examples of task-driven semantic coding on classification. From left to right, the images are the original image, semantic map, image coded with our RSC algorithm, images coded with equivalent semantic distortion and equivalent bits.}	
	\label{fig:class_img}
\end{figure*}
\section{Experiments}
\subsection{Experimental setting}
\subsubsection{Datasets}
As we employ the RL agent as the bit allocation predictor, we need a large quantity of training data. Therefore, we built a universe dataset for task-driven semantic coding, namely, the TSC dataset, which include three tasks: classification, detection, and segmentation. For classification, we selected 2,300 high-resolution images from the ImageNet \cite{deng2009imagenet} test set. Then, we resized them to 576x576. For detection and segmentation, we collect images from the TUD-Brussels pedestrian dataset \cite{wojek2009multi}, including 1,000 images. 

Then, all images are encoded with the H.265/HEVC reference software HM16.19\footnote{Available: \url{https://hevc.hhi.fraunhofer.de/svn/svn_HEVCSoftware/tags/HM-16.19/}}, while the QPs from 22 to 51 are applied for encoding. The interval of QPs for coding can be adjusted according to the semantic tasks. During encoding, we collected the bit cost of each CU from HM16.19. After encoding, we obtain the reconstructions of all images, which are sent to the vision task CNNs to obtain the corresponding semantic importance map. We use Grad-Cam for classification and Mask R-CNN for detection and segmentation. Finally, the TSC database is obtained, which is randomly divided into training (80\%) and test (20\%) sets. 
\begin{figure*}[htp]
	\begin{center}
		\includegraphics[width=1.0\textwidth]{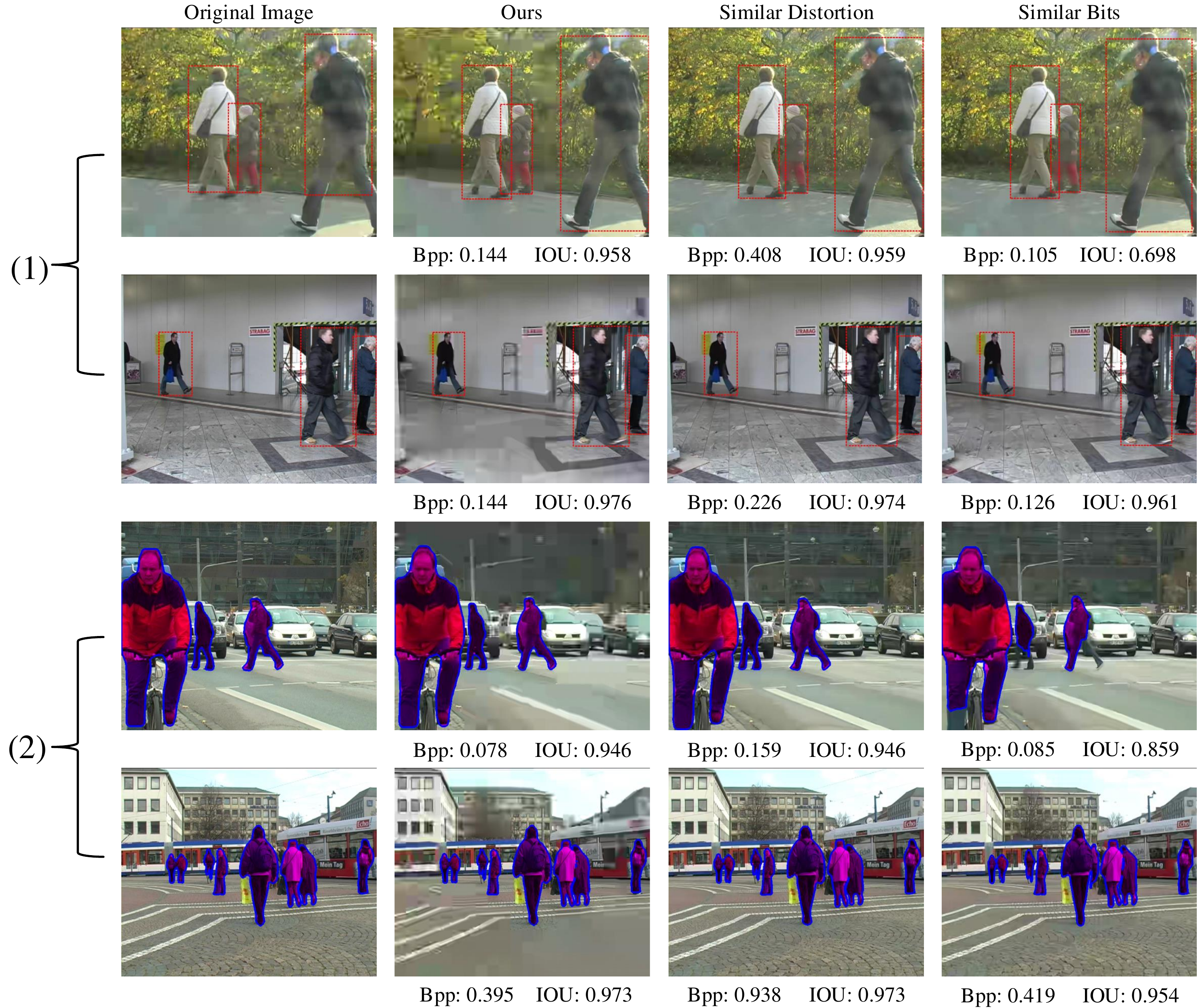}	
	\end{center}
	\caption{Examples of task-driven semantic coding on (1) detection and (2) segmentation. The first column is the original image before coding. The second column is coded with our proposed RSC algorithm. The third and fourth columns are coded with H.265/HEVC. The third column aims to compare the bits cost under the similar distortion, and the fourth column aims to compare the distortion under similar bits cost. Higher IOU means the better performance.}	
	\label{fig:seg}
\end{figure*}

\begin{figure*}[htp]
	\begin{center}
		\includegraphics[width=1.0\textwidth]{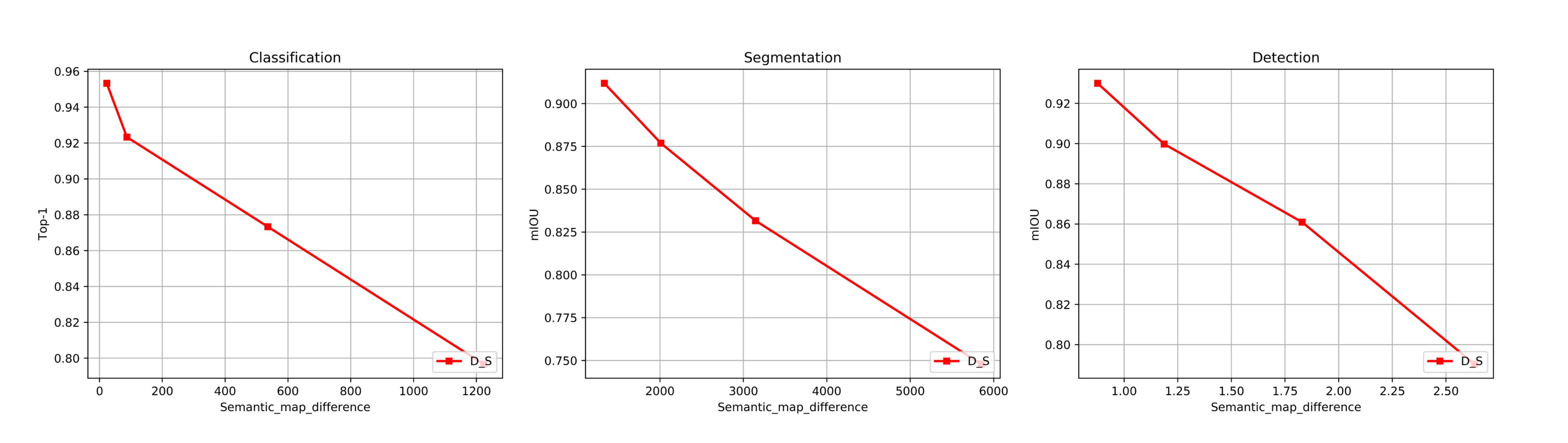}	
	\end{center}
 	\caption{The relationship between semantic fidelity and the semantic map difference for classification, segmentation and detection. Semantic fidelity is  negatively correlated with semantic map differences.}	
	\label{fig:semantic_analysis}
\end{figure*}  
\subsubsection{Training settings}
In this paper, we apply H.265/HEVC reference software HM16.19 as our codec. We compress our data with an all-intra main configuration. For DQN, the parameters are randomly initialized. We set the learning rate as 0.0001 and the discount factor as 0.9. The memory size is set to 50,000 and the mini-batch is set as 64. The target network parameters are updated every 300 steps. We utilize the TensorFlow framework to implement the whole model. It only takes 24 hours to train the DQN and obtain the best performance. The training process is as algorithm \ref{alg:Training}.
\subsubsection{Evaluation Standard}
To evaluate the effectiveness of our RSC, the average semantic fidelity and average bitrate are measured. To measure the task-driven semantic fidelity, we measure the top-5 accuracy for classification and the mean intersection of union (mIOU) for segmentation and detection.
  
\subsection{Compared with traditional codecs}
We compare our RSC with traditional codec H.265/HEVC. 
For H.265/HEVC, a larger quantization parameter (QP) means a higher compression ratio.  
According to experiments, we find that the top-5 accuracy for classification is 100\% when QP is lower than 27. Here, we select the QP as 27, 32, 37, 42, 47 for H.265/HEVC as our baseline. To validate the effectiveness of our task-driven semantic coding, we apply our algorithm in three tasks: classification, segmentation and detection. The process of RSC is shown in algorithm \ref{alg:process_RSC}.

\subsubsection{Classification}For classification, we set ${\alpha}_{s}$ in equation \ref{equation_16} as 0.005, 0.01, 0.025, 0.05, 0.25, 1, 2, 4 and 8, respectively. The RL agent focuses more on semantic fidelity when ${\alpha}_{s}$ becomes larger.  As shown in Fig. \ref{fig:result}(a), when ${\alpha}_{s}$ increases, the compression ratio decreases and semantic fidelity increases. Moreover, the proposed semantic coding method can obtain better classification accuracy compared with H.265/HEVC under the same bit cost. To better learn about how task-driven semantic coding works, we visualize the semantic map for classification as shown in Fig.\ref{fig:class_img}(b). The bright region represents the semantically important part of the image. According to the semantic map, the heads of rabbits, eyes of cats and cabin are essential for classification to make decisions. As shown in Fig. \ref{fig:class_img}(c), our methods can  better preserve the semantic fidelity at the above semantically important region. However, traditional codec wastes approximately 50\% bits on semantically unimportant regions.

\subsubsection{Segmentation and Detection}
 For different tasks, the measurements of semantic fidelity are different. Thus, ${\alpha}_{s}$ is also different. We empirically set the ${\alpha}_{s}$ as 0.005, 0.01, 0.025, 0.05, 0.1, 0.25, 0.5, 1, 2, 4, 8 and 16 for segmentation and set ${\alpha}_{s}$ as 0.01, 0.025, 0.05, 0.1, 0.25, 0.5, 1, 2, 4 and 8 for segmentation.  As shown in Fig. \ref{fig:result}(b) and (c), for segmentation, when QP in traditional coding is set as 27, the coding bitrate is approximately 6,400 kpbs/s. With the same mIOU, we can save approximately 50\% of the bits. For detection, when QP in traditional coding is set as 27, we can also save approximately 40\% of the bits, which is significant for data transmission and storage. The coded images are visualized as Fig. \ref{fig:seg}. 
 
 \subsubsection{BD-BR and BD-metric}
 In this section, we utilize Bjontegaard metric \cite{bjontegaard2001calculation} as our evaluation method for semantic coding scheme. Specifically, we compute the BD-BR and BD-metric \cite{pateux2007excel} for our RSC algorithm and utilize HM16.19 as our baseline. The BD-BR represents the average bit-rate reduction under the equivalent task-related accuracy, and the BD-metric represents the average task-related accuracy improvement under the equivalent bit-rate.
 The metrics for classification, segmentation and detection are Top-5 accuracy and mIOU, respectively. As shown in Table. \ref{tab:BDrate},  with the same semantic fidelity, our RSC algorithm can achieve 52.62\%, 51.01\% and 34.39\% bitrate savings on classification, segmentation and detection tasks, respectively, compared with HM16.19. Under the same bit cost, our RSC algorithm can improve the accuracy by 2.38\% on the classification task and the mIOU by 5.02\% and 3.11\%, respectively, on the segmentation task and detection task. 
 
 \begin{table}[htp]
 	\centering
 	\caption{BD-BR and BD-metric relative to the baseline in HM16.19.}
 	\label{tab:BDrate}
 	\setlength{\tabcolsep}{4mm}{
 		\begin{tabular}{c|c|c|c}
 			\hline
 			Tasks     & Classification & Segmentation & Detection \\ \hline
 			BD-BR     & -52.62\%       & -51.01\%     & -34.39\%  \\ \hline
 			BD-metric & 2.38\%         & 5.02\%       & 3.11\%    \\ \hline
 	\end{tabular}}
 \end{table}
 
 \subsection{Analysis of the semantic map}
 \label{semantic_map}
 In this section, we analyze the relationship between semantic map differences and semantic fidelity. To implement the semantic rate-distortion optimization, we represent the task-driven semantic fidelity with semantic map difference $\Delta {{M}_{s}}$, which leads the RL agent to balance the semantic fidelity and bit-rate. To validate its effectiveness, we allocate bits by selecting four QP values: 22, 27, 32, and 37. Then, we set these QPs as centers and randomly adapt the QP in the interval of plus or minus 5 for each CTU in one frame. The semantic importance is computed for the whole frame. We compute the classification accuracy and mIOU as our semantic fidelity for classification, segmentation and detection. The relationship of semantic map differences and semantic fidelity is shown in Fig. \ref{fig:semantic_analysis}. According to Fig. \ref{fig:semantic_analysis}, the semantic fidelity and semantic map difference are almost linear, which validates the effectiveness of our semantic map. Since the QP for each CTU is randomly set, the total semantic map differences of all CTUs are consistent with the semantic fidelity of the whole frame.

\subsection{Ablation Study}
\subsubsection{Comparison with handcrafted scheme}
To compare our method with the handcrafted scheme, after extracting the semantic map, we directly set the QPs for CUs of coding images instead of utilizing the RL agent to make a decision for classification. Specifically, we set larger QPs for semantically important regions and lower QPs for semantically unimportant regions. We set the QP from 22 to 51, which is the same as our action space. For a CTU of size 64x64, we first calculate the semantic importance by summing the corresponding semantic map as Eq. \ref{equation_s} and then normalize the S value.  
\begin{equation}
\label{equation_s}
S=\frac{\sum\limits_{x\in X}{\sum\limits_{y\in Y}{{{M}_{s}}}}}{XY},S=\frac{S}{{{S}_{MAX}}}.
\end{equation}

\begin{figure}[htp]
	\begin{center}
		\includegraphics[width=1.0\linewidth]{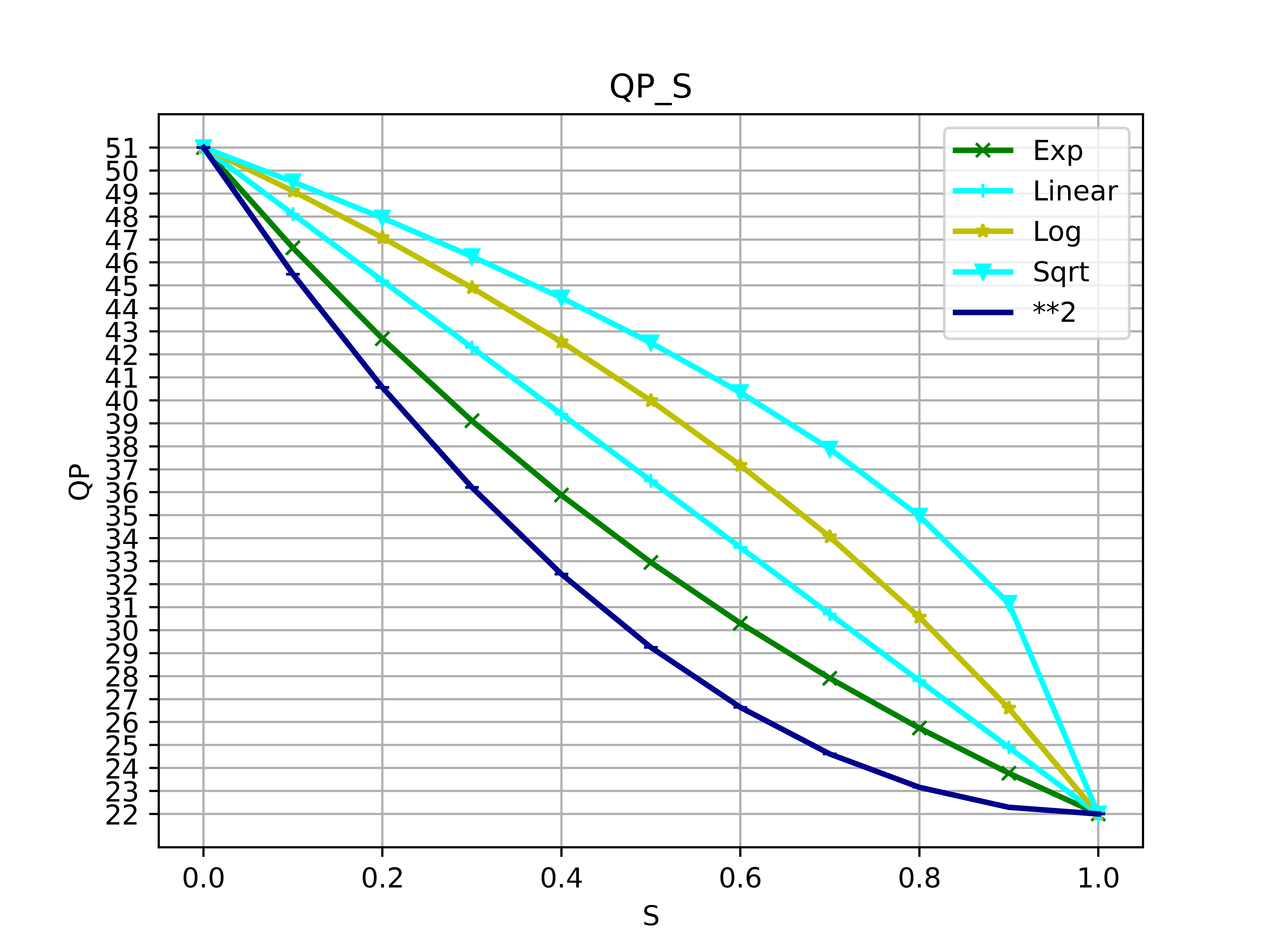}	
	\end{center}
	\caption{The method for calculating the QPs according to the semantic maps.}	
	\label{fig:qp_generate}
\end{figure}
\begin{figure}[htp]
	\begin{center}
		\includegraphics[width=1.0\linewidth]{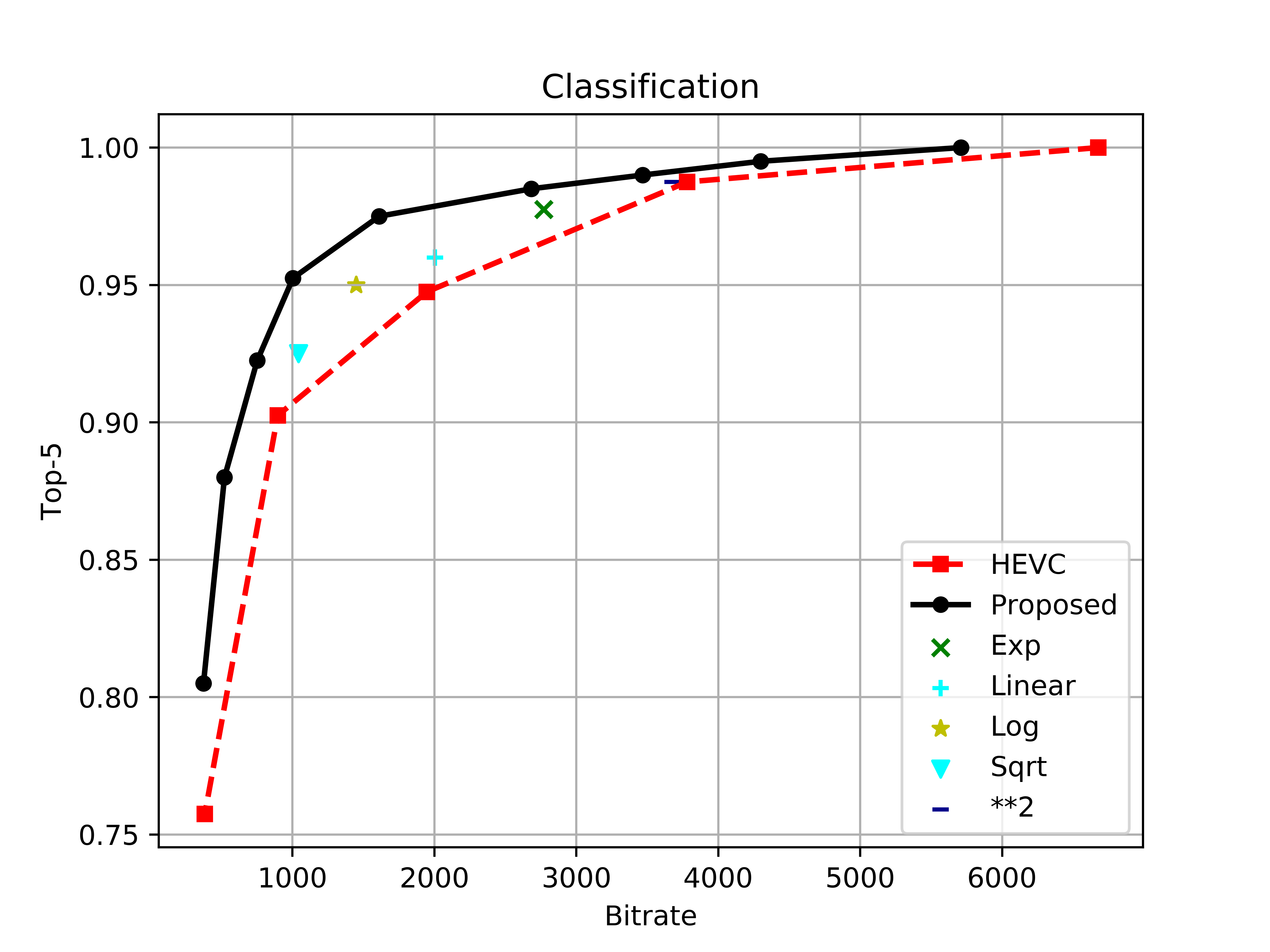}	
	\end{center}
	\caption{Comparison between our algorithm and handcrafted schemes.}	
	\label{fig:hand_draft}
\end{figure}

Then, we set the QP values with linear mapping or nonlinear mapping as shown in Fig. \ref{fig:qp_generate} according to the semantic importance $S$ since the relationship $S$ value and best QP value might be nonlinear. As shown in Fig. \ref{fig:hand_draft}, our algorithm is better than handcrafted schemes. There are two key reasons. First, handcrafted methods cannot adaptively adjust the QP value according to semantic importance because they cannot capture the best relationship between the QP value and semantic map. Second, it cannot balance the bitrate and semantic fidelity because the bitrate cannot be computed before coding. However, the RL agent can capture the above information with a learnable mechanism. With little training data, our method can adapt to the task and achieve stat-of-the-art semantic coding performance by utilizing semantic bit allocation. 

\subsubsection{Effects of global information for Q-network}
\begin{figure}[htp]
	\begin{center}
		\includegraphics[width=1.0\linewidth]{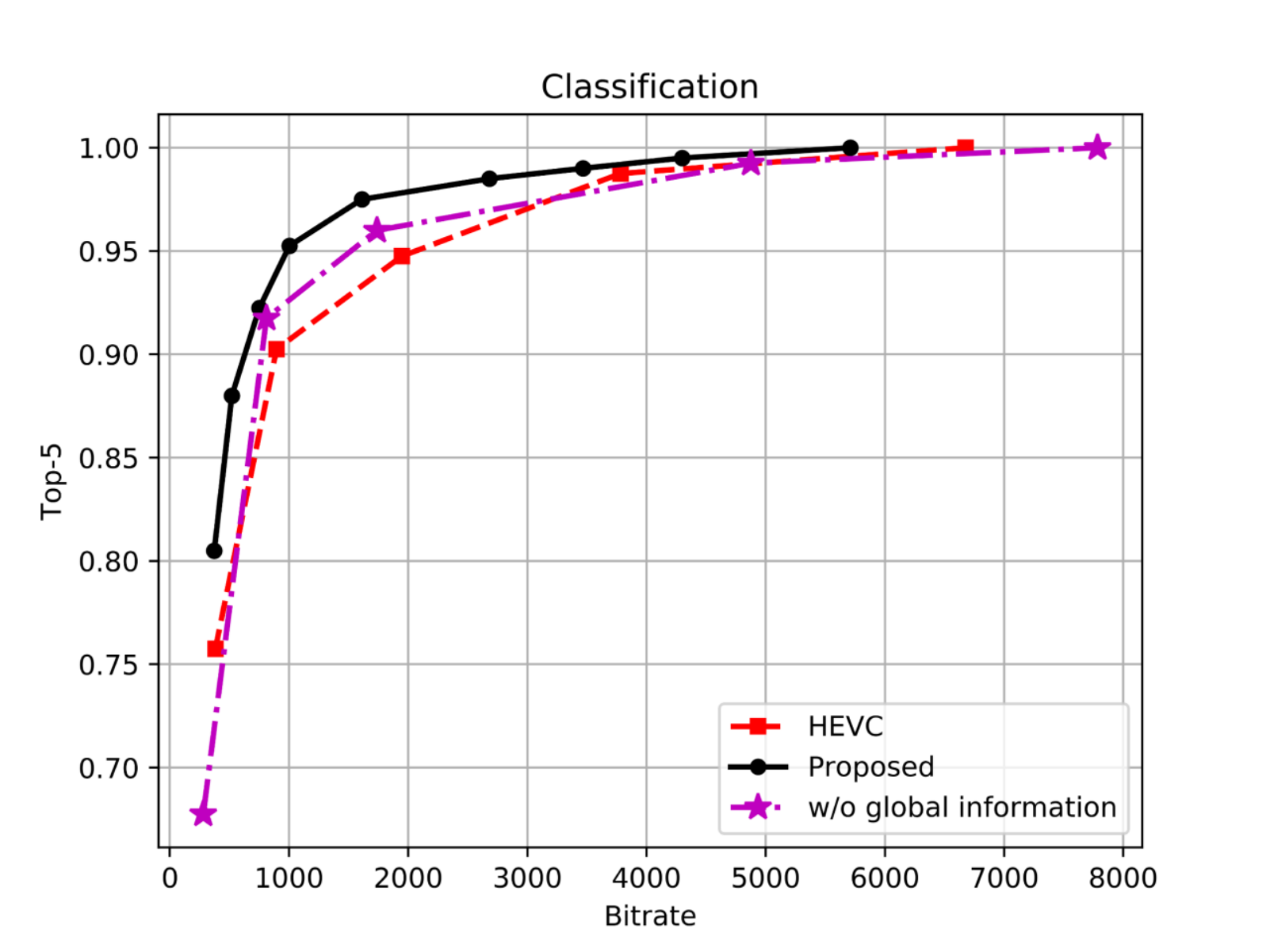}	
	\end{center}
	\caption{Effects of global information for Q-network.}	
	\label{fig:ab2}
\end{figure}

Since the quantization parameter (QP) determination of each block is not only associated with current block, but also associated with the state of whole coding image, our designed Q-network contains two branches, capturing global information and local information, respectively. The local information contains the current block and its semantic map, which is necessary to provide task-related information to the network. To validate the effectiveness of global information, we remove the branch used to capture the global information from our Q-network. And we select classification task to conduct experiments. The experimental results are as shown in Fig. \ref{fig:ab2}. The removing of global information causes obvious performance drop for our Q-network, which validate the effectiveness of our two-branches for Q-network. 

\subsection{Complexity Analysis}
In this section, we analyze the complexity of our algorithm. As shown in Table. \ref{complexity_analysis}, our algorithm does not change the decoding time. Moreover, the encoding time is almost the same as H.265/HEVC. The QP decision takes approximately 0.25s and semantic map generation only takes about 0.45s for one frame of size 576x576 when running with NVIDIA 2080Ti. It is efficient and effective to integrate our algorithm into intelligent media applications. The speed of QP decision and semantic map generation is not optimized in this work, which can be further optimized in future work. 

\begin{table}[htp]
	\centering
	\caption{Time complexity}
	\setlength{\tabcolsep}{4mm}{
	\begin{tabular}{c|l|c|c}
		\hline
		Run time                  & \multicolumn{2}{c|}{Encoder} & \multicolumn{1}{l}{Decoder} \\ \hline
		\multirow{3}{*}{Proposed} & RL agent  & 0.25s   & \multirow{3}{*}{0.036s}       \\ \cline{2-3}
		& Semantic map generation & \multicolumn{1}{l|}{0.45s}                        \\ \cline{2-3}
		& coding                & 1.735s  &                             \\ \hline
		HEVC                      & \multicolumn{2}{c|}{1.735s}     & 0.036 s                        \\ \hline
	\end{tabular}
}
\label{complexity_analysis}
\end{table}


\subsection{Comparisons of our RSC with its conference version}
In this section, we clearly clarify the difference between this paper and its conference version \cite{shi2020reinforced}, which can be summarized into following three parts. 
\subsubsection{Scalability and generalization}
\begin{figure}[htp]
	\begin{center}
		\includegraphics[width=1.0\linewidth]{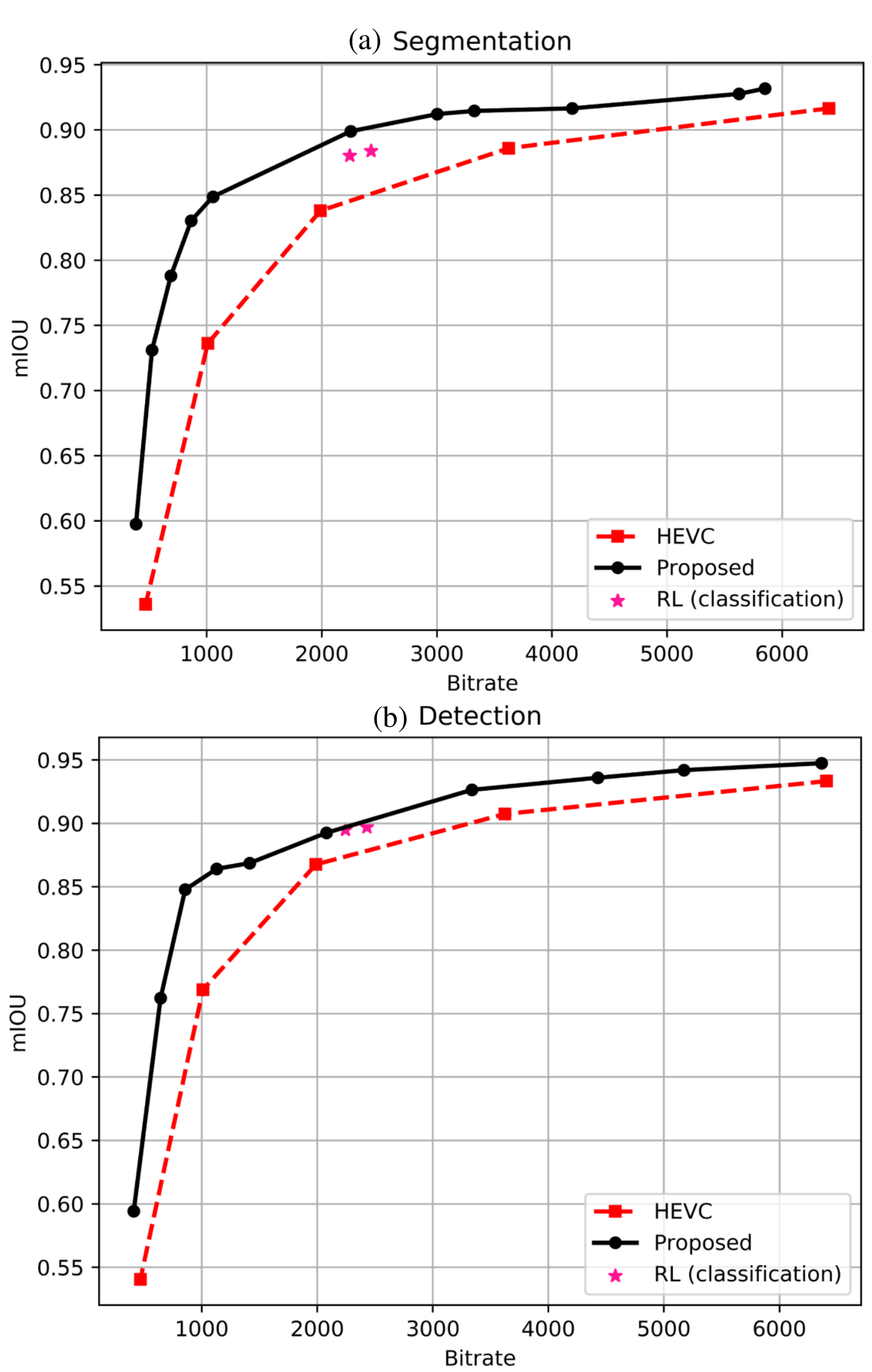}	
	\end{center}
	\caption{Scalability and generalization from classification task to segmentation and detection tasks. RL (classification) means that we utilize the RL agent trained only on classification task to determine the QPs for segmentation and detection.}	
	\label{fig:generalization}
\end{figure}

This paper builds a complete and general task-driven semantic coding framework by introducing a pixel-level semantic map. For the conference version   \cite{shi2020reinforced}, the different tasks have different semantic maps, which causes the inputs of RL network and reward definition are different. For example, the semantic map of detection task in the conference version \cite{shi2020reinforced} is the distribution of instance number and the semantic map of classification task is pixelwise importance map. Therefore, the conference version \cite{shi2020reinforced} lacks of scalability and generalization for different tasks. To overcome these issues, we introduce a pixel-wise semantic map to unify the different tasks in this journal paper. In this way, all tasks share the same RL architecture as shown in Fig. \ref{fig:DQN} and reward definition as Equ. \ref{equation_16}, which promises the scalability and generalization of our RSC (i.e, when meeting new task, the RL architecture and reward definition do not need any modification). To prove that, we utilize the RL agent trained only on classification task to determine the QPs for segmentation and detection tasks. As shown in the Fig.  \ref{fig:generalization}, our RL for classification can also achieve the considerate performance in detection and segmentation tasks, which validates the generalization of our semantic coding scheme in this paper.

\begin{figure}[htp]
	\begin{center}
		\includegraphics[width=1.0\linewidth]{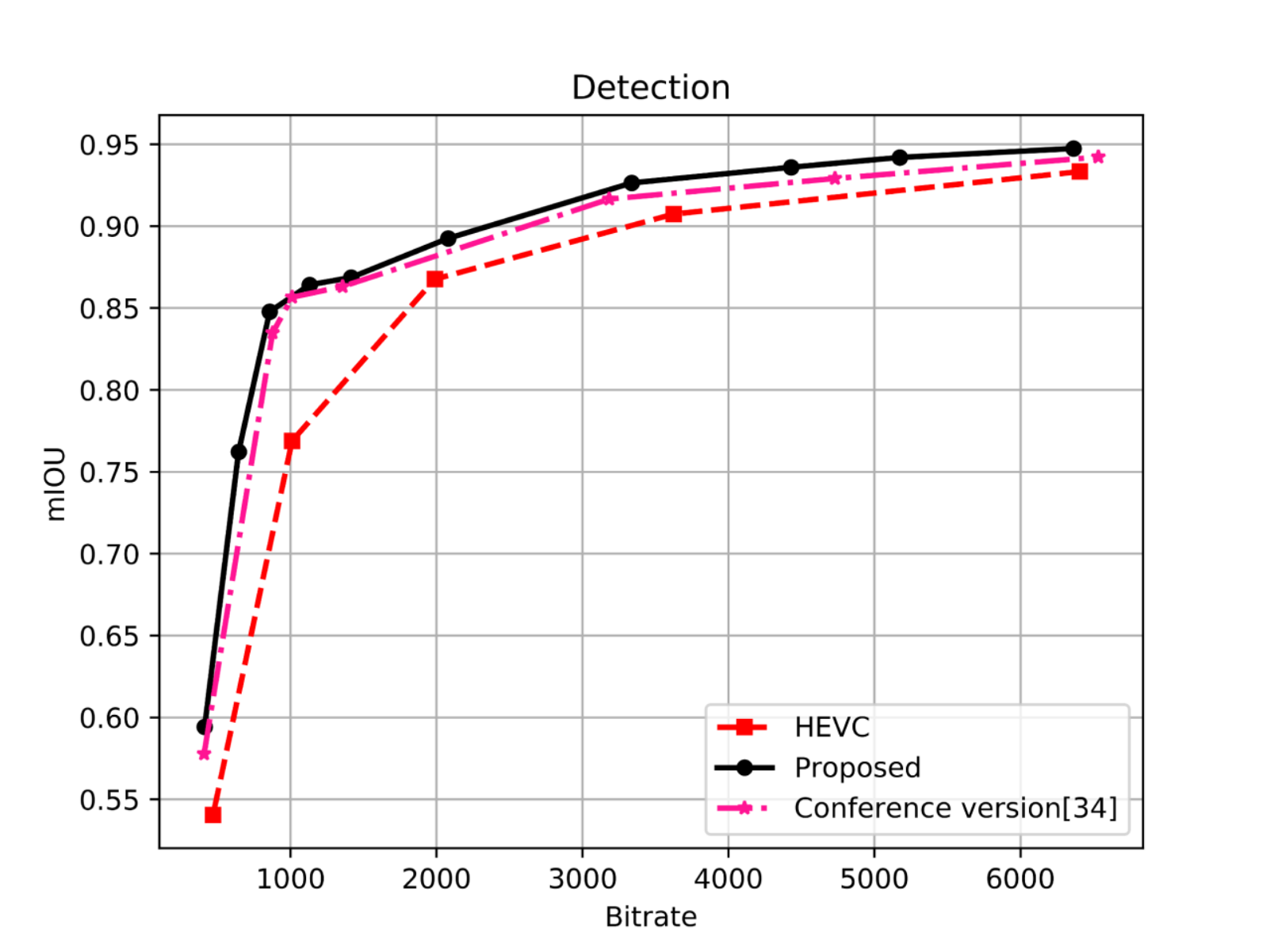}	
	\end{center}
	\caption{Comparison between our RSC with its conference version \cite{shi2020reinforced}.}
	\label{fig:ab_detection}
\end{figure}

\begin{figure}[htp]
	\begin{center}
		\includegraphics[width=1.0\linewidth]{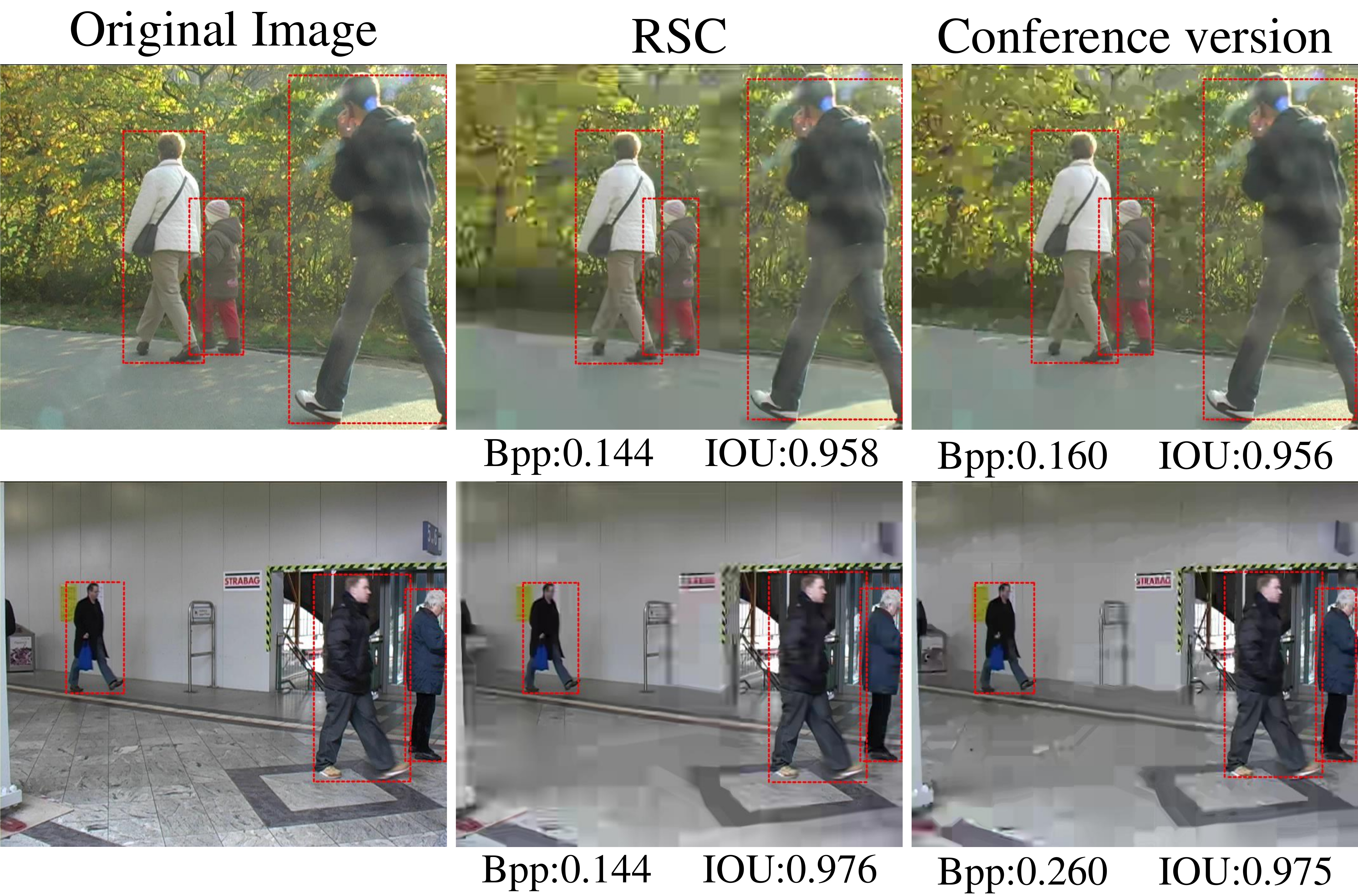}	
	\end{center}
	\caption{Comparison between our RSC with its conference version \cite{shi2020reinforced}.}
	\label{fig:com_with_conf}
\end{figure}

Moreover, the pixelwise semantic map can bring considerate improvement on detection task compared to the conference version \cite{shi2020reinforced}. The rate-distortion comparison on detection task of our RSC and its conference version \cite{shi2020reinforced} is shown in Fig. \ref{fig:ab_detection}. With the same semantic distortion, our RSC can save 8.81\% higher than its conference version \cite{shi2020reinforced}, which is shown in TABLE \ref{tab:BDrate_com}.  We also additionally add the subjective comparison of our RSC with its conference version [34] as shown in Fig. \ref{fig:com_with_conf}. 

 \begin{table}[htp]
 	\centering
 	\caption{BD-BR and BD-metric of our RSC and its conference version \cite{shi2020reinforced} with the baseline HM16.19.}
 	\label{tab:BDrate_com}
 	\setlength{\tabcolsep}{4mm}{
 		\begin{tabular}{c|c|c}
 			\hline
 			Tasks     & RSC &  Conference version \cite{shi2020reinforced} \\ \hline
 			BD-BR     &  -34.39\%  & -25.58\% \\ \hline
 			BD-metric &  3.11\%  & 2.53\%  \\ \hline
 	\end{tabular}}
 \end{table}

\subsubsection{Evaluation methods for task-driven semantic coding}
The conference version \cite{shi2020reinforced} of our RSC only works and is evaluated at one bitrate point. Moreover, we only compare our scheme with HEVC on QP 22 in the conference version, which is limited. To overcome these limitations, in this paper, we provide the formula \ref{equation_16} and change the bitrate of our RSC for coding by adjusting the parameter $\alpha$, that represents the importance degrees of semantic distortion. In this way, our RSC can be suitable for large range of bitrate. As shown in Fig. \ref{fig:result}, our RSC can achieves considerate performance improvement from low to high bitrate. To comprehensively evaluate our RSC, we also compute the BD-BR and BD-metric as shown in table \ref{tab:BDrate} to demonstrate the superiority of our RSC in this paper.

\subsubsection{Theory and experimental analysis}
In our conference version \cite{shi2020reinforced}, the reward definition is not clear and lacks of reasonable explanation. Besides, the effectiveness of importance map is not validated. Therefore, in this paper, we design the reward with detailed theory derivation as formula \ref{equation_16}. Moreover, we validate the effectiveness of our pixelwise semantic map to represent the task-driven semantic fidelity as shown in Fig. \ref{fig:semantic_analysis}. In addition, we provide the thorough experimental analysis for ablation study, complexity and stability.

\section{Conclusion}
In this paper, we first implement task-driven semantic coding for the traditional hybrid coding framework, which utilizes RL-based semantic bit allocation. Specifically, we design semantic maps for different tasks to extract the pixelwise semantic fidelity. Then, we utilize reinforcement learning (RL) to integrate the semantic fidelity metric into the in-loop optimization of semantic coding. Extensive experiments demonstrate the effectiveness of our algorithm. Our method can save  34.39\% to 52.62\% bits over the traditional coding framework with comparable semantic fidelity in different tasks such as classification, segmentation and detection. By designing the task-driven semantic map, our algorithm can be extended to other intelligent media applications easily without modifying the network for specific tasks.  In future work, we will consider to extend our scheme to support heterogeneous intelligent video tasks.

\appendices

\ifCLASSOPTIONcaptionsoff
  \newpage
\fi

\bibliographystyle{./bibtex/IEEEtran}
\bibliography{./bibtex/IEEEabrv,./myref}

\end{document}